\documentclass[
               twocolumn,
               noshowpacs,          
               nopreprintnumbers,     
               aps,                 
               prd,                 
               a4paper,             
               superscriptaddress,  
               nofootinbib,         
               tightenlines,        
               floats,floatfix      
               ]{revtex4}

\usepackage{amsmath, amsfonts, amsthm, amssymb, graphicx}
\usepackage{epstopdf}
 \usepackage{slashed}
 \usepackage{graphicx,epsfig}
\usepackage{amsmath}
\usepackage {amssymb}
\usepackage[utf8]{inputenc}
\usepackage{hyperref}
\usepackage[francais]{babel}
\usepackage[T1]{fontenc}

\newcommand{\be}{\begin{eqnarray}}
\newcommand{\ee}{\end{eqnarray}}
\newcommand{\bea}{\begin{eqnarray}}
\newcommand{\eea}{\end{eqnarray}}

\begin{document}

\title{3$d$ Quantum Gravity Partition Function at 3 Loops: a brute force computation}
\author{Mauricio Leston}
\affiliation{Instituto de Astronom\'{\i}a y F\'{\i}sica del Espacio IAFE-CONICET, Ciudad Universitaria, IAFE, 1428, Buenos Aires, Argentina.}
\affiliation{Departamento de F\'isica, Universidad de Buenos Aires FCEN-UBA and IFIBA-CONICET, Ciudad Universitaria, Pabell\'on 1, 1428, Buenos Aires, Argentina.}

\author{Andr\'es Goya}
\affiliation{Instituto de Astronom\'{\i}a y F\'{\i}sica del Espacio IAFE-CONICET, Ciudad Universitaria, IAFE, 1428, Buenos Aires, Argentina.}

\author{Guillem P\'erez-Nadal}
\affiliation{Departamento de F\'isica, Universidad de Buenos Aires FCEN-UBA and IFIBA-CONICET, Ciudad Universitaria, Pabell\'on 1, 1428, Buenos Aires, Argentina.}

\author{Mario Passaglia}
\affiliation{Departamento de F\'isica, Universidad de Buenos Aires FCEN-UBA and IFIBA-CONICET, Ciudad Universitaria, Pabell\'on 1, 1428, Buenos Aires, Argentina.}

\author{Gaston Giribet}
\affiliation{Department of Physics, New York University, 726 Broadway, New York, NY10003, USA.}



\begin{abstract}
The partition function of 3-dimensional quantum gravity has been argued to be 1-loop exact.
Here, we verify the vanishing of higher-orders in perturbation theory by explicit computation in the second-order, metric formulation at 3-loops. The number of 1-particle irreducible Feynman diagrams involving both gravitons and ghosts turns out to be 17. Using dimensional regularization, we solve all the diagrams. At 2-loops, we find that all such diagrams vanish separately after regularization. At 3-loops, in contrast, a series of remarkable cancellations between different diagrams takes place, with 9 diagrams beautifully conspiring to yield a vanishing result. Our techniques are suitable to be applied to higher loops as well as to similar computations in higher dimensions.
\end{abstract}

\maketitle

\section{Introduction} 
\label{sec:introduction}

The partition function of 3-dimensional quantum gravity has been shown to be 1-loop exact. In the case of the theory around Anti-de Sitter (AdS) space, this was proven by Maloney and Witten by the explicit computation of the sum over configurations \cite{MaloneyWitten}, and this turns out to be consistent with the following argument: In 3-dimensional gravity around AdS, in addition to the classical action, the partition function receives contributions of states that are Virasoro descendants of the background geometry. This follows from the analysis of the asymptotic symmetries in AdS performed by Brown and Henneaux \cite{BrownHenneaux}. Such contributions, often referred to as boundary gravitons, organize themselves as Virasoro descendants, and their logarithm, being independent of the Planck length, is identified as the 1-loop contribution to the effective action. This insight led Maloney and Witten to argue that, without further contributions, the full gravity partition function around AdS turns out to be 1-loop exact, with the only non-vanishing contributions being the classical action and the Virasoro character, cf. \cite{Witten, MaloneyWitten, Giombi}. In the case of the theory with zero cosmological constant the 1-loop exactness of the 3$d$ gravity partition function was discussed by Witten in an earlier paper \cite{Witten0}, where a computation in the first-order formulation was given, cf. \cite{CS}. Witten argued that the perturbative expansion in the theory must terminate at 1-loop. More recently, the authors of \cite{Glenn} found that the 1-loop determinant computation of the gravity partition function reproduces the character of the Bondi-Metzner-Sachs (BMS) group, namely the group of asymptotic diffeomorphisms preserving the boundary conditions of Minkowski space at null infinity \cite{Glenn2, Glenn4, Glenn5}. This led to the conclusion that, as it happens in AdS, the partition function of 3$d$ Einstein gravity in flat space is also 1-loop exact, with the full effective action being given by the classical contribution plus a group character, cf. \cite{Glenn6, Glenn7, Mau, Bianca1}. However, it still remains to be seen how the 1-loop exactness of the partition function manifests itself in the second-order, metric formulation, especially because previous calculations, while heuristically convincing, are not conclusive: the path integral computation performed in \cite{Witten0} for flat space included degenerate metrics  in the configuration space, and the computation for AdS in \cite{MaloneyWitten} relies on the Brown-Henneaux prescription for the configuration space and leads to a final result whose interpretation still remains unclear. As the authors of \cite{Glenn} stated, it would be interesting to verify the 1-loop exactness of the 3$d$ gravity from a direct gravitational computation. This is exactly the computation we will address in this paper: we will compute the partition function of 3$d$ gravity partition function around flat space in the metric formalism at third order in perturbation theory. That is to say, we will perform an explicit field theory computation of the gravitational effective action at 2- and 3-loops.

The paper is organized as follows: In section \ref{sec:II}, we present the tools that will equip us for the perturbative computation. We write down the gravity action in a convenient form, discuss the Faddeev-Popov gauge fixing terms and the action for the ghost fields; then, we write the vertices and the propagators for the ghost and the graviton; all these ingredients suffice to derive the Feynman rules. In section \ref{sec:III}, we compute all the Feynman diagrams. At 2-loops, we find that all connected diagrams vanish separately after dimensional regularization, in agreement with previous computations in the literature. At 3-loops, in contrast, a series of remarkable cancellations between different diagrams takes place, with 9 1-particle irreducible (1PI) diagrams beautifully conspiring to yield a vanishing result. As the steps we follow in this work can be applied to higher loops and also adapted to higher dimensions, we briefly comment on that at the end of section \ref{sec:III}. Section \ref{sec:IV} contains our conclusions. 

\section{Perturbation theory}
\label{sec:II}

The Einstein-Hilbert gravitational action is
\begin{equation}
    \label{EHaction}
    S_{\rm EH} = -\dfrac{2}{\kappa^2}\int_{M_d} d^d x \sqrt{|g|}\, R \, +\, B_{\partial {M_d}}\,,
\end{equation}
where $|g|$ is the determinant of the metric and $R$ is the scalar curvature on the $d$-dimensional manifold $M_d$. $B_{\partial {M_d}}$ stands for boundary terms that render the variational principle well-posed. The coupling constant $\kappa^2=32\pi \ell_P^{d-2}$ gives the Planck length, $\ell_P$, in $d$ spacetime dimensions. The quantity $\kappa^2\hbar$ organizes the loop expansion. Hereafter, we set $\hbar=1$ and keep track of powers of $\kappa$.

A convenient way to rewrite the Einstein-Hilbert action is the following
\begin{equation}
\begin{split}
    S_{\rm EH} = -\frac{1}{2\kappa^2}&\int_{M_d} d^d x \sqrt{|g|} \, g^{mn} g^{ab} g^{rs} \Big(  \partial_{m}{g_{ab}} \partial_{n}{g_{rs}} -\\ & 
     \partial_{m}{g_{ar}} \partial_{n}{g_{bs}}  + 2 \partial_{m}{g_{br}} \partial_{a}{g_{ns}} - 2 \partial_{m}{g_{na}} \partial_{b}{g_{rs}}   \Big) \,\nonumber
\end{split}
\end{equation}
where we are excluding boundary terms \cite{LandauLifshitz.Vol2}. 

We will perform an expansion of the gravitational field around a background metric $\bar{g}_{ab}$, namely
\begin{equation}
    \label{pert}
    g_{ab} = \bar{g}_{ab} + \kappa \, h_{ab} \, .
\end{equation}
In our case, the background metric will be that of flat space, {\it i.e.} $\bar{g}_{ab}\equiv \eta_{ab}$; nevertheless, for completeness, let us write some formulae in full generality; this will allow to show that the techniques we employ are applicable to computations around other maximally symmetric solutions. In the action above one replaces the perturbation around $\bar{g}_{ab}$ and obtains an infinite series in $h_{ab}$; this obviously follows from expanding $\sqrt{|g|}$, $g_{ab}$ and $g^{ab}$ in terms of $h_{mn}$. The first terms of the expansion come from ${|g|}^{\frac 12} = {|\Bar{g}|^{\frac 12}} ( 1 + \frac 12 \kappa \bar{g}^{mn}h_{mn} + \ldots )$ and $g^{ab} = \Bar{g}^{ab} -\kappa \bar{g}^{am} \bar{g}^{bn} h_{mn} + \ldots$, where the ellipsis stand for subleading (higher order) contributions. At order $\mathcal{O}(\kappa^0h^2)$ we have the canonically normalized quadratic kinematic operator; at order $\mathcal{O}(\kappa h^3)$, the 3-graviton vertex; at order  $\mathcal{O}(\kappa^2 h^4)$, the 4-graviton vertex, and so on and so forth. At 3-loops, there are be contributions up to order $\mathcal{O}(\kappa^4 h^6)$; see diagram $D_4^{(3)}$ below.

Einstein-Hilbert action has to be supplemented with gauge-fixing terms. The piece of the full action that implements the gauge fixing reads
\begin{equation}
\label{gaugefix}
    S_{\rm gf} = \int_{M_d} d^dx \sqrt{|g|} \, f^{m} \bar{g}_{mn} f^{n} \,,
\end{equation}
where the function $f^m$ on the background metric $\bar{g}_{mn}$ with a perturbation $ h_{mn}$ is given by
\begin{equation}
\label{harmonicgauge}
    f^{m} = \left( \Bar{g}^{lm} \bar{\nabla}^{n} - \frac12 \bar{g}^{ln} \bar{\nabla}^{m} \right) h_{nl} \, ;
\end{equation}
$\bar{\nabla}$ is the covariant derivative compatible with the background metric $\bar{g}$; namely $\bar{\nabla}_{a}\bar{g}_{cb}=0$.

The action for the ghost field $c^s$ and the anti-ghost field $\Bar{c}^l$ is
\begin{equation}
    \label{ghostaction}
    S_{\rm gh} =  \int_{M_d} d^d x \sqrt{|g|} \, \Bar{c}^{m} \,\dfrac{\delta f_{m}}{\delta h_{rs}} \, \mathcal{L}_{c} g_{rs} \,,
\end{equation}
where ${\mathcal L}_{c} g_{rs}$ is a Lie derivative of the full metric $g_{ab}$ with the respect to the ghost field $c^l$
\begin{equation}
    \label{Liedrivghost}
    {\mathcal L}_{c} g_{rs} = 2 \bar{g}_{l(s}\bar{\nabla}_{r)}{c^l} + c^l\Bar{\nabla}_{l}{h_{rs}} + 2 h_{l(s}\Bar{\nabla}_{r)}{c^l} \,.
\end{equation}
Then, up to a total derivative, the ghost action takes the form
\begin{eqnarray}
    S_{\rm gh} &= 
      \int_{M_d} d^d x \sqrt{|g|} \,
    \Big[  \bar{g}_{ls} \Bar{c}^{s} \Bar{\nabla}^2{c^l} + \Bar{c}^{r}\Bar{\nabla}_{l}\Bar{\nabla}_{r}{c^l} - \Bar{c}^{m}\Bar{\nabla}_{m}\Bar{\nabla}_{l}{c^l}  \nonumber \\ 
    &- \kappa \Big(  \Bar{\nabla}^{r}{\Bar{c}^{s}} \Bar{\nabla}_{l}{h_{sr}} c^l + \Bar{\nabla}^{r}{\Bar{c}^{s}} h_{ls} \Bar{\nabla}_r{c^l}   
    +\Bar{\nabla}^{s}{\Bar{c}^{r}} h_{ls} \Bar{\nabla}_{r}{c^l}
    \nonumber \\ &
    - \frac12 \Bar{\nabla}_{m}{\Bar{c}^m}\Bar{\nabla}_{l}{{h}_{r}^{\, r}}\,c^l - \Bar{\nabla}_{m}{\Bar{c}^{m}} h_{ls} \Bar{\nabla}^{s}{c^l} \Big)\Big]    \label{ghostlagrangiannexp}
\end{eqnarray}
These formulae are consistent with the ones in the literature \cite{Grisaru, Sagnotti, deWitt}. 

Hereafter, we restrict the discussion to flat space. This amounts to replacing $\bar{g}_{ab}\to \eta_{ab}$ and $\bar{\nabla}_m\to \partial_m$ in the expressions above. In flat space, the ghost propagator is
\begin{equation}   \label{ghostprop}
    \Delta_{\rm (gh)}^{ab}[k] = -i \dfrac{\eta^{ab}}{k^2}\, .
\end{equation}
with $k$ being the momentum. 

The Einstein-Hilbert action takes the form
\begin{equation}
\begin{split}
\label{EHactionLandauh}
    S_{\rm EH} =&-\frac 12  \int_{M_{d}} d^d x \sqrt{|g|} \, g^{mn} g^{ab} g^{rs} \Big(   \partial_{m}{h_{ab}} \partial_{n}{h_{rs}} -\\ &  \partial_{m}{h_{ar}} \partial_{n}{h_{bs}}  + 2 \partial_{m}{h_{br}} \partial_{a}{h_{ns}} - 2 \partial_{m}{h_{na}} \partial_{b}{h_{rs}}   \Big)\, , \,\nonumber
\end{split}
\end{equation}
which can further be expanded in powers of $\kappa h^{ab}$. This yields the graviton vertices and the graviton propagator
\begin{equation}
\label{gravitonprop}
    \Delta_{\rm (gr)}^{mnab}[k] = \dfrac{i}{2k^2} \left( \eta^{ma}\eta^{nb} + \eta^{mb}\eta^{na} - \dfrac{2}{d-2}\eta^{mn}\eta^{ab} \right)\, 
\end{equation}

Propagators (\ref{ghostprop}) and (\ref{gravitonprop}) are written Minkowski signature. Here, however, we will work in the Euclidean formalism. This amounts to carefully collect relative signs of vertices and propagators in the diagrams. In momentum space $k^{\mu}=(k^0,k^1,k^2)$, we perform the Wick rotation $k^0\to ik^3$, with $M_3$ being now locally equivalent to $ \mathbb{R}^3$ with Euclidean signature. Since we are interested in computing the partition function at finite temperature, we consider the periodic Euclidean time direction, $M_3= \mathbb{R}^2\times S^1_{\beta}$, which demands the momentum $k^3$ along the thermal cycle to be quantized; namely $k^3=2\pi n/\beta $, with $n\in \mathbb{Z}$ and $\beta \in \mathbb{R}_{>0}$ being the inverse of the temperature. 
Nevertheless, we will abuse of notation and write the formal sum $\int d^dk_l\, (.)$ to refer to the integration measure on the $l^{\rm th}$ $d$-momentum, while keeping in mind the sum over discrete values for the component $k_l^3$.

In the next section, we will apply the ingredients presented above to the computation of the gravitational effective action at 2-loops and 3-loops.

\section{3-loop partition function}
\label{sec:III}

\subsection{Partition function}

The statement that the 3$d$ gravitational partition function is 1-loop exact can easily be translated into a statement about the effective action. Consider the expansion of partition function 
\begin{equation}
\log Z \, = \, \sum_{n=0}^\infty {\hbar}^{n-1} S_{\rm eff}^{(n)}
\end{equation}
with $S_{\rm eff}^{(n)}$ being the $n^{\text th}$ order contribution to the effective action, with $S_{\rm eff}^{(n)}\sim\mathcal{O}(\kappa^{2(n-1)})$; $S_{\rm eff}^{(0)}$ and $S_{\rm eff}^{(1)}$ being given by the classical action and the 1-loop determinant, respectively. Then, 1-loop exactness is equivalent to assert that $S_{\rm eff}^{(n)}=0$ to all order $n>1$ in perturbation theory -- or, more precisely, that higher orders in perturbation theory only contribute to the renormalization of the parameters appearing in the semiclassical theory \cite{Giombi}--. Here, we will prove that $S_{\rm eff}^{(2)}$ and $S_{\rm eff}^{(3)}$ are actually zero.

\subsection{2-loops}
\label{sec:2loops}

\begin{figure}[ht]
\centering
\includegraphics[scale=0.40]{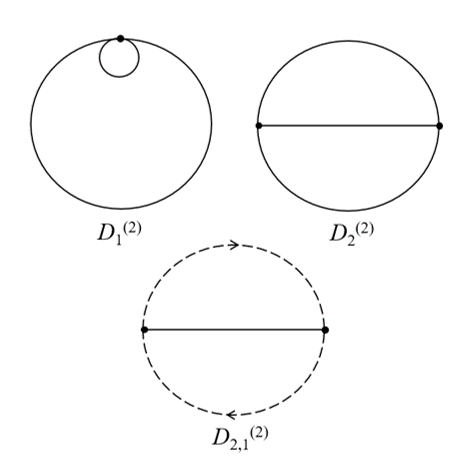}
\caption{2-loop 1PI diagrams.} \label{Fig1}
\end{figure}

\begin{figure}[ht]
\centering
\includegraphics[scale=0.42]{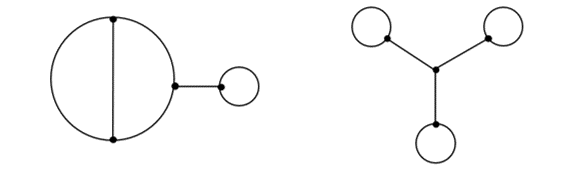}
\caption{Two examples of the 6 reducible 3-loop diagrams} \label{Figl2}
\end{figure}

At 2-loops, there are only three 1PI diagrams, which are depicted in Figure \ref{Fig1}. 
While we have to solve all connected diagrams, all those that are reducible (Figure \ref{Figl2}) vanish by dimensional regularization. This means that we have to focus on the 1PI. We have to solve the Feynman type integrals corresponding to each of such diagrams and perform dimensional regularization. We introduce the notation $d_i^{(\ell )}=[D_i^{(\ell )}]$ to denote the result of the calculation of the $i^{\rm th}$, $\ell $-loop diagram $D_i^{(\ell )}$. The symbol $[X]$ refers to the value of the quantity $X$ after dimensional regularization. It turns out that the three diagrams in Figure \ref{Fig1} also vanish after dimensional regularization. That is to say, $d_1^{(2)}=[D_1^{(2)}]=0$, $d_2^{(2)}=[D_2^{(2)}]=0$, and $d_{2,1}^{(2)}=[D_{2,1}^{(2)}]=0$. Therefore, we find
\begin{equation}
S_{\rm eff}^{(2)}\, = \, 0\, ,
\end{equation}
in full agreement with the argument in \cite{Glenn} and the calculations in \cite{Odintsov1, Odintsov2}. In the next subsection we will prove that this results holds at 3-loops.

\subsection{3-loops}
\label{sec:3loops}

At 3-loops the story is much more involved. The difficulty resides, not only in the fact that the number of 1PI diagrams is notably larger, but also in that many diagrams do not vanish after dimensional regularization, and, therefore, nontrivial cancellations have to occur for the effective action to be zero. Some of the diagrams have terms proportional to the integral
\begin{equation}
I= \kappa^{4}\int \prod_{i=1}^3d^{3}k_i \, \, \frac{(k_1\cdot k_2)\, (k_1\cdot k_3)}{k_1^2k_2^2k_3^3(k_1+k_2+k_3)^2} \, .\label{LaIntegral}
\end{equation}
This integral is relatively simple to treat at zero temperature, but it becomes more subtle at finite temperature: we have to be reminded of the fact that the formal integral over momenta in (\ref{LaIntegral}) actually comprises the sum on the component $k_i^3=2\pi n_i/\beta $ with $n_i\in\mathbb{Z}$, $i=1,2,3$ and $\beta\in\mathbb{R}_{>0}$. That is to say, $I$ is defined as an infinite sum over integers $n_1, \, n_2, \, n_3$; more precisely,
\begin{eqnarray}
I&=& {32\pi^5\kappa^4\beta^{-5}} \sum_{n_1, n_2, n_3}\int \prod_{i=1}^3d^{2}\vec{k}_i \,  \prod_{j\neq 1}^{3}(\vec{k}_j\cdot \vec{k}_1 + n_j\, n_1) \nonumber\\
&&
\Big[\big| \sum_{t=1}^{3} \vec{k}_t\big|^2
+ \big(\sum_{t=1}^{3}n_t\big)^2\, \Big]^{-1}
\,\prod_{l=1}^{3}\big(|\vec{k}_l|^2+n_l^2\, \big)^{-1} \, \label{LaIntegralbeta}
\end{eqnarray}
where $\vec{k}_i=\frac{\beta}{2\pi}(k_i^1,k_i^2)$, so that $k_i=\frac{2\pi}{\beta}(\vec{k}_i,n_i^3)$. This integral is similar to those appearing in other 3-loop computations at finite temperature \cite{Integrales1, Integrales2, Integrales3}. As these integrals appear in several diagrams at 3-loops, cancellations among different diagrams are actually possible -- and, as we will see, they actually take place. 

\begin{figure}[ht]
\centering
\includegraphics[scale=0.40]{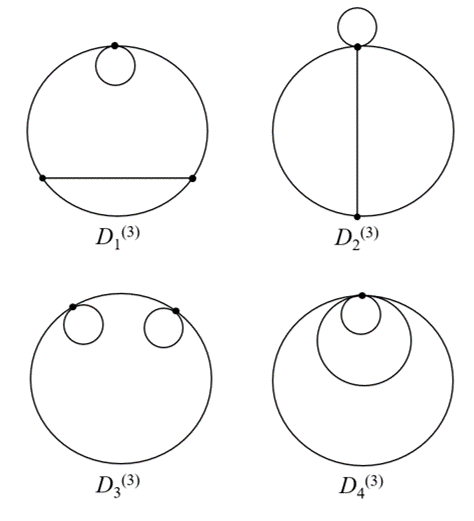}
\caption{3-loop diagrams that vanish after regularization.} \label{Fig2}
\end{figure}

Other divergent integrals that may potentially appear in the 3-loop 1PI diagrams are of the form 
\begin{equation}
J_i= \kappa^{4}\int \prod_{l=1}^3d^{3}k_l \, \, j_i(k_1, k_2,  k_3)\, ,\label{LaSIntegralES}
\end{equation}
with integrands 
\begin{equation}
j_1=\frac{\mathcal{O}(k^2)\, }{k_1^2k_2^2k_3^2} \,,  \ \ 
j_2=\frac{\mathcal{O}(k^4)\, }{k_1^2k_2^2k_3^2(k_2+k_3)^2} \,,  \ \
j_3=\frac{\mathcal{O}(k^4)\, }{k_1^4k_2^2k_3^2} \, ;  \, \nonumber
\end{equation}
however, the latter cancel in $d=3$, either by dimensional regularization and/or the presence of factors $(d-3)$. This is one of the reasons why the diagrams appearing in Figure \ref{Fig2} do not effectively contribute; {\it i.e.} $d_1^{(3)}=d_2^{(3)}=d_3^{(3)}=d_4^{(3)}=0$. 

\begin{figure}[ht]
\centering
\includegraphics[scale=0.40]{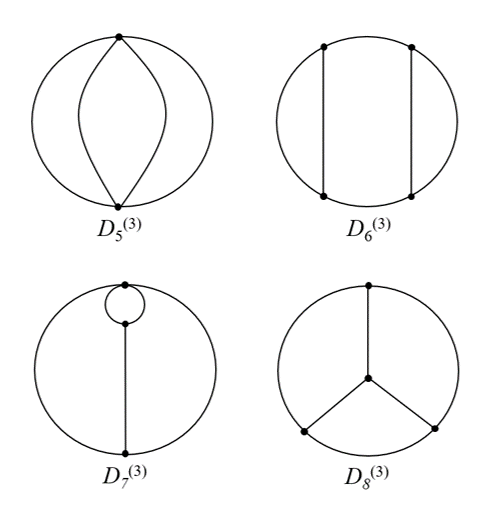}
\caption{3-loop graviton diagrams.} \label{Fig3}
\end{figure}

Next, we focus on the diagrams shown in Figure \ref{Fig3}, whose propagator contributions make their dimensional regularization analysis more involved.
In order to solve the diagrams $D_5^{(3)}, D_6^{(3)}, D_7^{(3)},\, D_8^{(3)}$, first one needs to compute the multiplicity factor associated to each of them. If $g_i^{(3)}$ denotes the multiplicity factor corresponding to the diagram $D_i^{(3)}$, then combinatorics yields 
\begin{equation}
g_5^{(3)}=\frac{1}{24}\, , \ \ \ g_6^{(3)}=\frac{1}{16}\, , \ \ \ g_7^{(3)}=\frac{1}{8}\, , \ \ \ g_8^{(3)}=\frac{1}{24}\, . \nonumber
\end{equation}
After multiplying these factors by the result of each diagram obtained after dimensional regularization, we find
\begin{eqnarray}
d_5^{(3)}=  \frac{45}{16} I\, ,  \  \ 
d_6^{(3)}=  I \, , \ \   d_7^{(3)}=-\frac{61}{16} I \, , \ \  
d_8^{(3)}=  \frac 32  I \,  \nonumber
\end{eqnarray}
where $I$ is the integral given in (\ref{LaIntegralbeta}). The evaluation of these diagrams, especially the Melon $D_5^{(3)}$ and the Benz $D_8^{(3)}$, is lengthy and requires precision. 

\begin{figure}[ht]
\centering
\includegraphics[scale=0.40]{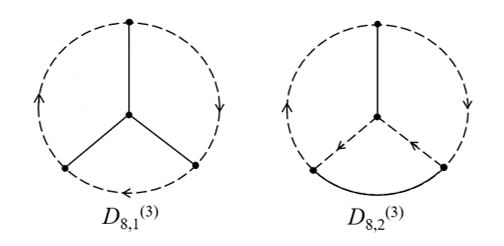}
\caption{3-loop diagrams with ghosts fields; cf. \cite{deWitt}.} \label{Fig4}
\end{figure}

Now, let us consider the diagrams with ghost field contributions. We begin with the ghostly Benz diagrams shown in Figure \ref{Fig4}. 
It turns out that, after dimensional regularization, these diagrams also result proportional to (\ref{LaIntegralbeta}). Concretely, they yield
\begin{eqnarray}
d_{8,1}^{(3)}=  \frac{11}{4} I\, , \ \ \ \ 
d_{8,2}^{(3)}=  -\frac{13}{8} I \,  ; \nonumber
\end{eqnarray}
having multiplicity factors $g_{8,1}^{(3)}=1/3$ and $g_{8,2}^{(3)}=1/4$, respectively. The different signs between $d_{8,1}^{(3)}$ and $d_{8,2}^{(3)}$ follows from the number of ghost propagators in each diagram.

\begin{figure}[ht]
\centering
\includegraphics[scale=0.40]{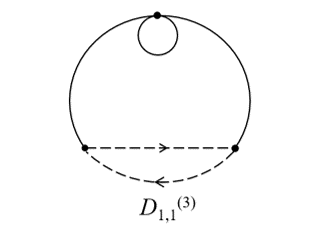}
\caption{3-loop diagrams with ghosts fields.} \label{Fig5}
\end{figure}

There are also diagrams with ghost contributions that vanish directly by dimensional regularization; the diagram $D_{1,1}^{(3)}$ shown in Figure \ref{Fig5} is of that sort, {\it i.e.} $d_{1,1}^{(3)}=0$. 

\begin{figure}[ht]
\centering
\includegraphics[scale=0.40]{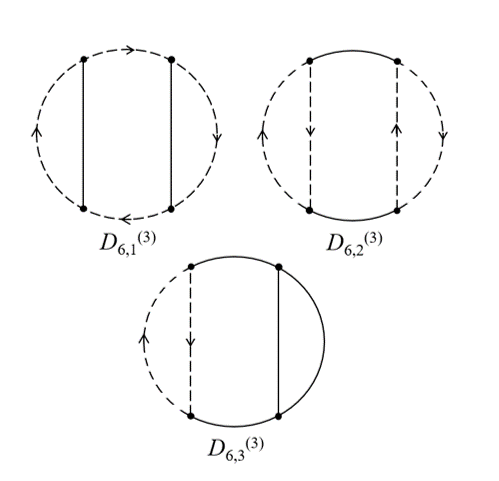}
\caption{3-loop diagrams with ghosts fields.} \label{Fig6}
\end{figure}

The non-vanishing 3-loop diagrams with ghost contributions are depicted in Figure \ref{Fig6}. Their evaluation is lengthy but can be done systematically. It yields
\begin{eqnarray}
d_{6,1}^{(3)}= \, -\frac{1}{2}\, I\, , \ \ \ 
d_{6,2}^{(3)}= \, -\frac{3}{8}\, I \, , \ \ \ 
d_{6,3}^{(3)}= \, -\frac 74 \, I \, . \nonumber
\end{eqnarray}
The multiplicity factors of these diagrams are $g_{6,1}^{(3)}= 1/2$, $g_{6,2}^{(3)}= 1/4$, and $g_{6,3}^{(3)}=1/4$, respectively.

Finally, putting all this together, we find that the 3-loop contribution to the gravitational effective action reduces to an expression proportional to $I$, with a coefficient proportional to
\begin{equation}
d_{5}^{(3)}+ d_{6}^{(3)}+d_{6,1}^{(3)}+d_{6,2}^{(3)}+d_{6,3}^{(3)}+d_{7}^{(3)}+d_{8}^{(3)}+d_{8,1}^{(3)}+d_{8,2}^{(3)} =  0 \nonumber
\end{equation}
Therefore, we finally find
\begin{equation}
S_{\rm eff}^{(3)}\, = \, 0\, .\label{3L}
\end{equation}
in full agreement with \cite{Glenn}. This result follows from a notable cancellation among different diagrams; a cancellation that decomposes as follows
\begin{equation}
\Big(\frac{45}{16}+1-\frac{61}{16}+\frac 32 +\frac{11}{4}-\frac{13}{8}-\frac 12 -\frac 38 -\frac 74\Big) \, I\, = \, 0\nonumber  
\end{equation}
with each term in the sum coming from a different diagram. Notice that in this cancellation there are also partial cancellations, {\it e.g.} $d_{5}^{(3)}+ d_{6}^{(3)}=-d_{6,1}^{(3)}$. It would be desirable to understand if there is a precise reason for the cancellation of different subsets of diagrams, and thus gain intuition that could serve us for calculations at higher loops. The fact that all diagrams cancel is remarkable, as it might have happened that the graviton and ghost contributions did not completely cancel at higher loops; see the discussion in \cite{Glenn} about this point; see also \cite{Guardini} and references thereof. It would also be interesting to compare the cancellation we observed here and the analogous remarkable cancellations that often occur in scattering amplitudes calculation, cf. \cite{Parke}.  

\subsection{Higher dimensions}

In order to further study the origin of the cancellation expressed by equation (\ref{3L}), we find illustrative to discuss the computation in $d$ dimensions. Besides, this allows us to emphasize that our techniques are well suited to be extended to arbitrary dimension $d\geq 3$, something that to some extent is obvious as we have been working with dimensional regularization. 

It can be shown that the $d$-dimensional analog to (\ref{3L}) turns out to be proportional to sum of terms of the form
\begin{equation}
{\kappa^4} \, \frac{(d-3)}{(d-2)^2}\, P_i(d)\, I_i\label{HHH}
\end{equation}
where $I_i$ stand for the the $d$-dimensional extension of integral (\ref{LaIntegralbeta}) and for other integrals that appear in $d>3$, and $P_i(d)$ are polynomials. This manifestly shows that the cancellation in (\ref{3L}) only happens for $d=3$. It is worth mentioning that, in addition to the eight diagrams depicted in Figures \ref{Fig3}, \ref{Fig4} and \ref{Fig6}, other diagrams also contribute to (\ref{HHH}) when $d> 3$. In $d$ dimensions there may be additional integrals to be solved as we have checked that several contributions throughout the computation vanish due to factors of $(d-3)$. In spite of all that, our calculation can well be extended to $d\geq 3$.

\subsection{Higher loops}

\begin{figure}[ht]
\centering
\includegraphics[scale=0.40]{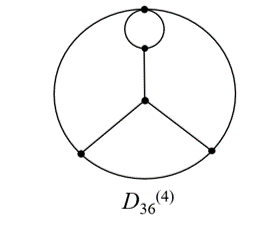}
\caption{Example of a 4-loop graviton diagram.} \label{Fig7}
\end{figure}

Before concluding, a few words about higher loops: While our computation can in principle be extended to higher-loops, the calculation becomes rapidly unmanageable due to the increasing number of 1PI diagrams. At 4-loops the number of diagrams happens to be, roughly, one order of magnitude larger than the number of them at 3-loops. Graviton vertices of order $\mathcal{O}(\kappa^6h^8)$ start to contribute and the plethora of graphs becomes unwieldy. Still, one can say a few things about the 4-loop 1PI contributions; for example, that there are many diagrams that vanish after dimensional regularization, while other diagrams, {\it e.g.} the one depicted in Figure \ref{Fig7}, happen to be more involved. Therefore, one also expects cancellations similar to the one we obtained in (\ref{3L}) to take place at 4-loops.

\section{Conclusions}
\label{sec:IV}

In this paper, we computed the 3-dimensional gravitational effective action in flat spacetime at 2- and 3-loops in the second-order, metric formalism. We showed that the result vanishes, in agreement with \cite{Witten0, MaloneyWitten, Giombi, Glenn}. The computation amounted to handle the ghost and graviton contributions, and solve the integrals associated to all connected Feynman diagrams. The calculation, being lengthy and requiring precision, demanded the implementation of a systematic procedure. By dealing with dimensional regularization, we solved all the connected 2- and 3-loop diagrams. At 2-loops, we found that all the diagrams vanish in any number of dimensions. This led us to explore the next order in the loop expansion. At 3 loops, there are 14 1PI diagrams, 9 of them surviving after carefully performing dimensional regularization. Crucial to the final result was a remarkable cancellation among the latter, with the 9 diagrams conspiring to yield a vanishing result in $d=3$. Consequently, our computation turns out to be a non-trivial check of the 1-loop exactness of the 3$d$ partition function \cite{Witten0}. In other words, we have provided a consistency check of the result presented in \cite{Glenn}, where the authors argued that the quantum corrections to 3$d$ gravity partition function around flat space is fully determined by a 1-loop determinant that reproduces the character of the BMS group.

\subsection*{Acknowledgments}
The authors thank Glenn Barnich, David Blanco, Stephen Carlip, Alan Garbarz for discussions. M.L. thanks the CCPP at NYU for the hospitality during his stay, where part of this work was done. The computation presented in this paper partially resorted to FeynCalc \cite{R1,R2,R3}. The computational resources used in this work were provided in part by the HPC center DIRAC, funded by IFIBA (UBA-CONICET) and part of SNCAD-MinCyT initiative, Argentina. This work has been partially supported by grants PIP-(2022)-11220210100685CO, PIP-(2022)-11220210100225CO, PICT-(2021)-GRFTI-00644, PICT-2020-SERIEA-00164.

\[ \]

\providecommand{\href}[2]{#2}\begingroup\raggedright\endgroup
\end{document}